\documentclass[
aps,%
12pt,%
final,%
notitlepage,%
oneside,%
onecolumn,%
nobibnotes,%
nofootinbib,% In the current version of REVTeX, when this option is on,
superscriptaddress,%
noshowpacs,%
centertags]%
{revtex4}
\RequirePackage{graphicx}

\def\refitem#1{\relax}

\begin{document}
\title{Domain Growth and Ordering Kinetics in Dense Quark Matter}

\author{\firstname{Awaneesh} \surname{Singh}}
%\email{awaneesh11@gmail.com}
\affiliation{ School of Physical Sciences, Jawaharlal Nehru University, New Delhi 110 067,
 India}

\author{\firstname{Sanjay} \surname{Puri}}
%\email{puri@mail.jnu.ac.in}
\affiliation{ School of Physical Sciences, Jawaharlal Nehru University, New Delhi 110 067,
India}

\author{\firstname{Hiranmaya} \surname{Mishra}}
\email{hm@prl.res.in}
\affiliation{Theory Division, Physical Research Laboratory, Ahmedabad 380 009, India}

\begin{abstract}
The kinetics of chiral transitions in quark matter is studied in a two flavor Nambu-Jona-Lasinio model. We focus on the phase ordering dynamics subsequent to a temperature quench from the massless quark phase to the massive quark phase. We study the dynamics by considering a phenomenological model (Ginzburg-Landau free-energy functional). The morphology of the ordering system is characterized by the scaling of the order-parameter correlation function.
\end{abstract}

\maketitle

Heavy-ion collision experiments at high energies produce hot and dense strongly-interacting matter and provide the opportunity to explore the phase diagram of QCD \cite{cpod}. It is worthwhile to stress that, in a phase transition process, information about which equilibrium phase has lowest free energy is not sufficient to discuss all possible structures that the system can have. One has to understand the kinetics of the process by which the phase ordering or disordering proceeds and the nature of nonequilibrium structures that the system must go through on its way to reach equilibrium \cite{pw09,aj94}. In the present work, we study kinetics of chiral transitions in hot and dense matter \cite{dima}.

We use the two-flavor Nambu-Jona-Lasinio (NJL) model to study the chiral symmetry breaking in QCD \cite{klevansky}. The expression for the thermodynamic potential is obtained in the mean-field approximation:           
\begin{eqnarray}
\tilde\Omega(M,\beta,\mu) &=& \;-\frac{12}{(2\pi)^3\beta}
\displaystyle\int \! d\vec{k}\;\Big\{ \ln\left[1+e^{-\beta\left( \sqrt{k^2+M^2} 
- \mu \right)}\right] 
 + \ln\left[1+e^{-\beta\left( \sqrt{k^2+M^2} + \mu \right)}\right] \Big\}   
\nonumber\\& -&\frac{12}{(2\pi)^3}\displaystyle\int \! d\vec{k} \; 
\left(\sqrt{\vec{k}^2+M^2}-|\vec k|\right)+ \frac{M^2}{4G}.
\label{tomega}
\end{eqnarray}
Here, we have taken vanishing current quark mass, and introduce the constituent mass $M=-2g\rho_s$ with $\rho_s=\langle\bar\psi\psi\rangle$ being the scalar density and $g=G[1+1/(4N_c)]$. The details of the mean-field approximation are reported elsewhere \cite{awaneesh1}.

\begin{figure}[!tb]
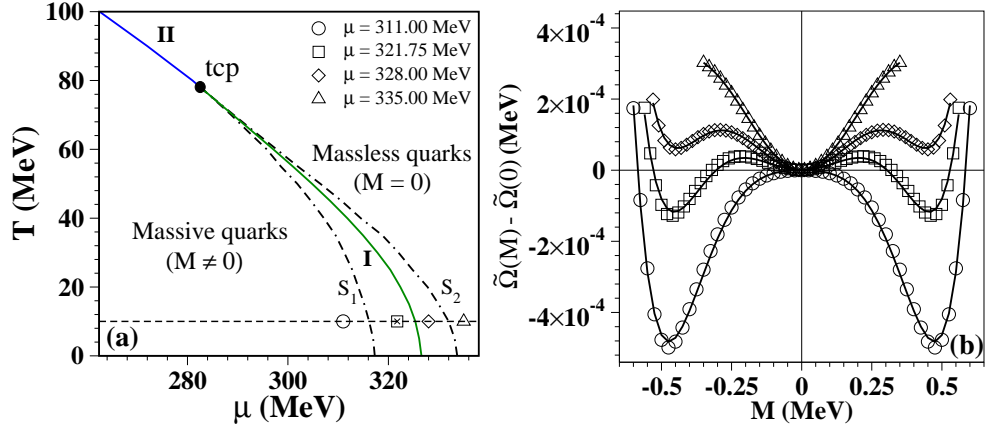

\centering
\begin{tabular}{c c }
\includegraphics[width=0.38\textwidth]{fig1a.eps}&
\includegraphics[width=0.4\textwidth]{fig1b.eps}\\
\end{tabular}
\vspace{-0.35cm}
\caption[tiny]{\footnotesize (a) Phase diagram of the Nambu-Jona-Lasinio (NJL) model in the ($\mu, T$)-plane 
for zero current quark mass. A line of first-order transitions (I, green online) meets a line of second-order transitions (II, blue online) at the tricritical point (tcp). We have $(\mu_{tcp}, T_{tcp}) \simeq (282.58, 78)$ MeV. The dot-dashed lines $S_1$ and $S_2$ denote the spinodals or metastability limits for the first-order transitions. The open symbols denote 4 combinations of $\left(\mu, T\right)$, chosen to represent qualitatively different shapes of the NJL potential. The cross denotes the point at which we quench the system for $b<0$. (b) Plot of $\tilde\Omega\left(M, \beta, \mu \right)$ from Eq.~(\ref{tomega}) as a function of $M$. The ($\mu, T$)-values are marked in (a). The solid lines superposed on the potentials correspond to the GL potential in Eq.~(\ref{p6}).}
\label{fig1}
\end{figure}

The phase diagram resulting from Eq.~(\ref{tomega}) is shown in Fig.~\ref{fig1}(a). For the parameters of the NJL model, we have taken the three-momentum cutoff $\Lambda=653.30$ MeV, and coupling $G=5.0163\times 10^{-6}$ $\mathrm{MeV^{-2}}$ \cite{askawa}. With these parameters, the vacuum mass of quarks is $M\simeq 312$ MeV. At $T=0$, a first-order transition takes place at $\mu \simeq 326.321$ MeV. For $\mu=0$, a second-order transition takes place at $T \simeq 190$ MeV. The first-order line (I) meets the second-order line (II) at the tricritical point $(\mu_{tcp},T_{tcp})
\simeq(282.58, 78)$ MeV. 

Close to the phase boundary, the potential in Eq.~(\ref{tomega}) may be expanded as a Ginzburg-Landau (GL) potential in the order parameter $M$: 
\begin{equation}
\tilde\Omega\left(M \right)= \tilde\Omega\left(0 \right) + \frac{a}{2}M^2 + \frac{b}{4}M^4 + \frac{d}{6}M^6 + O(M^8),
\label{p6}
\end{equation}
correct upto logarithmic corrections \cite{glfree}. In the following, we consider the expansion of potential $\tilde\Omega\left(M \right)$ upto the $M^6$-term. The first two coefficients in Eq.~(\ref{p6}) can then be exactly obtained by comparison with Eq.~(\ref{tomega}) while the coefficients $b$ and $d$ are obtained by fitting the GL potential with the potential Eq.~(\ref{tomega}) for a given $T$ and $\mu$. For stability, we require $d> 0$.

The extrema of the potential in Eq.~(\ref{p6}) are determined by the gap equation $\tilde\Omega^\prime(M)=0$ with solution $ M=0$, and $M_{\pm}^2= (-b\pm \sqrt{b^2 -4ad})/(2d)$. For $b>0$, the transition is second-order, with the stationary points  $M=0$ (for $a>0$) or $M=0$, $\pm M_+$ (for $a<0$). For $a<0$, the preferred equilibrium state is the one with massive quarks. For $b<0$, the solutions of the gap equation are as follows: (i) $M=0$ for $a>b^2/(4d)$, (ii) $M=0$, $\pm M_+$, $\pm M_-$ for $b^2/(4d)>a>0$, and (iii) $M=0$, $\pm M_+$ for $a<0$. A first-order transition takes place at $a_c=3b^2/(16d)$ with the order parameter jumping discontinuously from $M=0$ to $M=\pm M_+=\pm (3|b|/4d)^{1/2}$.

The evolution of the system is described by the time-dependent Ginzburg-Landau (TDGL) equation: 
\begin{equation}
\frac{\partial}{\partial t}M\left(\vec{r},t\right)= -\Gamma 
\frac{\delta \Omega\left[M \right]}{ \delta M}+\theta\left(\vec{r},t\right), 
\label{ke}
\end{equation}
which models the over-damped relaxational dynamics of $M(\vec{r},t)$ to the minimum of $\Omega\left[M \right]$ \cite{hohenrev}. Here, $\Gamma$ is the inverse damping coefficient, and $\theta(\vec{r},t)$ is the noise term satisfying the fluctuation-dissipation relation: $\left\langle\theta\left(\vec{r},t\right) \right\rangle = 0$ and $\left\langle \theta(\vec{r'},t')\theta(\vec{r''},t'') \right\rangle = 2\Gamma T\delta(\vec{r'}-\vec{r''})\delta\left(t'-t''\right)$. We use the natural scales of order parameter, space and time to introduce dimensionless variables: $M=M_0M'$ ($M_0=\sqrt{|a|/|b|}$); $\vec{r}=\xi\vec{r'}$ ($\xi=\sqrt{K/|a|}$); $t=\tau t'$ $[\tau=(\Gamma|a|)^{-1}]$; $\theta=(\Gamma|a|^{3/2}T^{1/2}/|b|^{1/2})~\theta'$. Dropping the primes, we obtain the dimensionless TDGL equation:
\begin{eqnarray}
\frac{\partial}{\partial t}M\left(\vec{r},t\right)=-\mathrm{sgn}
\left(a\right)M - \mathrm{sgn}\left(b\right)M^3-\lambda M^5 + \nabla^2 M +\theta\left(\vec{r},t\right),
\label{ke2}
\end{eqnarray}
where $\lambda=|a|d/b^2 >0$.

\begin{figure}[!tb]
\centering
\includegraphics[width=0.6\textwidth]{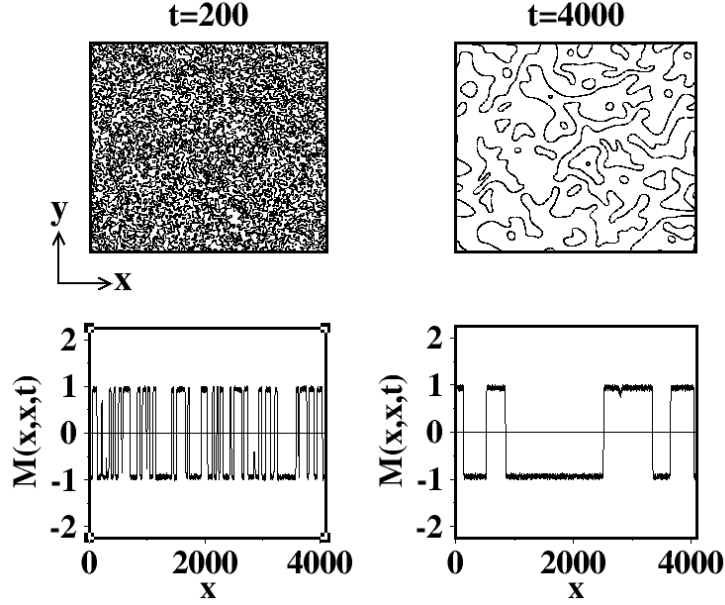}
\vspace{-0.5cm}
\caption[tiny]{\footnotesize Interface evolution after temperature quench through second-order line (II) in Figs.~\ref{fig1}(a). The snapshots at the top show the interfaces ($M=0$) at $t=200$, $4000$ (in units of $\tau$). They were obtained by numerically solving Eq.~(\ref{ke2}) as described in the text with $a<0$, $b>0$, $\lambda=0.14$. The noise amplitude was $\epsilon=0.008$. The frames at the bottom show the variation of the order parameter along the diagonal.}
\label{fig2}
\end{figure}

First, we study the ordering dynamics for $b>0$ corresponding to a quenching through the second order transition. We solve Eq.~(\ref{ke2}) numerically using an Euler-discretization scheme with an isotropic Laplacian implemented on a $d$=2 lattice of size $N^2$ ($N=4096$), with periodic boundary conditions in all directions. The details of the simulation can be found elsewhere \cite{awaneesh1}. In Fig.~\ref{fig2}, we show the evolution of a disordered initial condition for Eq.~(\ref{ke2}) with $b>0$ and $a<0$ -- corresponding to a temperature quench through the second-order line (II) in Fig.~\ref{fig1}(a). The initial state consists of small-amplitude thermal fluctuations about the massless phase $M=0$. The system rapidly evolves into domains of the massive phase with $M\simeq M_+$ and $M\simeq-M_+$. The interfaces between these domains correspond to $M=0$ -- their evolution is shown in the snapshots (frames at the top) of Fig.~\ref{fig2}. The frames at the bottom show the order parameter variation of the snapshots at the top. 

The domains have a characteristic length scale $L(t)$, which grows with time. The growth process is analogous to coarsening in the TDGL equation with an $M^4$-potential \cite{pw09, aj94}. The order-parameter correlation function $C(r,t)$ shows \emph{dynamical scaling} $C(r,t)=f(r/L)$. The scaling function  $f(x)=(2/\pi)\sin^{-1}(e^{-x^2})$ has been calculated by Ohta \emph{et al.} (OJK) \cite{ojk82} in the context of an ordering ferromagnet. Further, the domain scale obeys the Allen-Cahn (AC) growth law, $L(t)\sim t^{1/2}$ \cite{aj94}. The same growth law has also been obtained via a closed time path formalism of relativistic finite-temperature field theory applied to the NJL model \cite{das}.

\begin{figure}[!tb]
\centering
\includegraphics[width=0.6\textwidth]{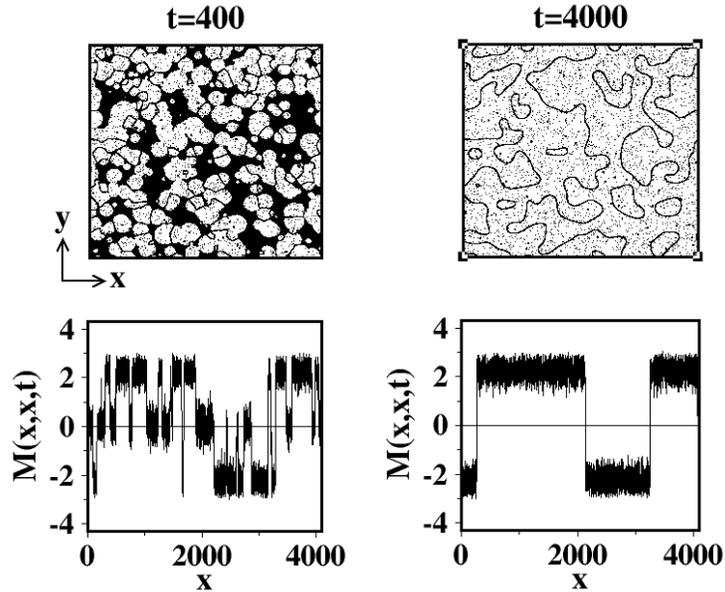}
\vspace{-0.5cm}
\caption[tiny]{\footnotesize Interface evolution after temperature quench through first-order line (I) in Figs.~\ref{fig1}(a). The snapshots at the top show the interfaces ($M=0$) at $t=400$, $4000$. They were obtained by solving Eq.~(\ref{ke2}) with $b<0$, $a_c>a>0$ and $\lambda=0.14$. The frames at the bottom show the variation of the order parameter along the diagonal. Notice that the metastable patches ($M=0$) at $t=400$ are absent at later times.}
\label{fig3}
\end{figure}

Next, we consider the case with $b<0$ . In this case, a first-order chiral transition occurs for $a<a_c=3b^2/(16d)$ (or $\lambda<\lambda_c=3/16)$. For $a<0$, the potential has a double-well structure and the ordering dynamics is equivalent to $M^4$-theory, i.e., the domain growth scenario is similar to Fig.~\ref{fig2}. We focus on a quench from the disordered state (with $M=0$) to $0<a<a_c$ or $0<\lambda< \lambda_c$, corresponding to a quench between the first-order line (I) and $\mathrm{S_1}$ in Fig.~\ref{fig1}(a) -- the corresponding points are denoted by crosses. The massless state is now a metastable state of the $M^6$-potential. The chiral transition proceeds via the nucleation and growth of droplets of the preferred phase ($M=\pm M_+$). The nucleation results from large fluctuations in the initial condition or thermal fluctuations during the evolution. In Fig.~\ref{fig3}, we show the nucleation and growth process. At early times ($t=400$), the system is primarily in the $M=0$ phase with small droplets of the preferred phase. These droplets grow in time and coalesce into domains. 

In the late stages of growth, there is no memory of the nucleation dynamics which characterized growth during the early stages. Let us note that a large dissipation (i.e., $\Gamma^{-1}$) will make the equilibration time larger. If this becomes larger than the life time of the fire ball in heavy ion collision, the system may linger in the symmetry restored phase longer even when the temperature has already decreased below $T_c$. This means the $T_c$ calculated based on equilibrium thermodynamic models could be higher than the value that shows up in the experiments \cite{dima, awaneesh1}.

To summarize: we have studied the kinetics of chiral phase transitions in QCD. In terms of the quark degrees of freedom, the phase diagram is obtained using the NJL model. An equivalent coarse-grained description is obtained from an $M^6$-Ginzburg-Landau (GL) free energy. The chiral kinetics is modeled via the TDGL equation, and we consider the overdamped case. We study the ordering dynamics resulting from a sudden temperature quench through the first-order (I) or second-order (II) transition lines in Figs.~\ref{fig1}(a). For shallow quenches through II and deep quenches through I, the massless phase is spontaneously unstable and evolves to the massive phase via spinodal decomposition. For shallow quenches through I, the massless phase is metastable and the phase transition proceeds via the nucleation and growth of droplets of the massive phase. The merger of these droplets results in late-stage domain growth analogous to that for the unstable case. In all cases, the asymptotic growth process exhibits dynamical scaling, and the growth law is $L(t)\sim t^{1/2}$. Given the dynamical universality of the processes involved, our results are of much wider applicability than the simple NJL Hamiltonian considered here.

\newpage
%%%%%%%%%%%%%%%%%%%%%%%%%%%%%%%%%%%%%%%%%%%%%%%%%%%%%%%%%%%%%%

%%%%%%%%%%%%%%%%%%%%%%%%%%%%%%%%%%%%%%%%%%%%%%%%%%%%%%%%%%%%%%

%%%%%%%%%%%%%%%%%%%%%%%%%%%%%%%%%%%%%%%%%%%%%%%%%%%%%%%%

%%%%%%%%%%%%%%%%%%%%%%%%%%%%%%%%%%%%%%%%%%%%%%%%%%%%%%%%
\end{document}